\documentclass[12pt,a4paper]{article}

\usepackage{amsmath}
\usepackage{amssymb}
\usepackage{amsfonts}
\usepackage{graphicx}
\usepackage{subfig}
\usepackage{caption}
\usepackage{dcolumn}
\usepackage{bm}
\usepackage[
colorlinks,linkcolor=blue,citecolor=magenta]{hyperref}
\usepackage[top=3cm,right=2cm,bottom=2.5cm,left=2cm]{geometry}
\begin{document}
\title{Topological invariants of the Ryu-Takayanagi ($RT$) surface used to observe holographic superconductor phase transition}

\author{Fatemeh Lalehgani Dezaki$^{a}$\thanks{f.lalehgani@ph.iut.ac.ir}, Behrouz Mirza$^{a}$\thanks{b.mirza@cc.iut.ac.ir}, Marzieh Moradzadeh$^{a}$\thanks{Marzieh.moradzadeh@ph.iut.ac.ir},\\ Zeinab Sherkatghanad$^{a}$\thanks{sherkat.elham@gmail.com}\\
	\it $^{a}$Department of Physics, Isfahan University of Technology, \\Isfahan, 84156-83111, Iran}

\maketitle
\begin{abstract}
We study the phase transitions in the metal/superconductor system using topological invariants of the Ryu-Takayanagi ($RT$) surface and the volume enclosed by the $RT$ surface in the Lifshitz black hole background. It is shown that these topological invariant quantities identify not only the phase transition but also its order. According to these findings a discontinuity slope is observed at the critical points for these invariant quantities that correspond to the second order of phase transition. These topological invariants provide a clearer illustration of  the superconductor phase transition than do the holographic entanglement entropy and the holographic complexity. Also, the backreaction parameter, $k$, is found to have an important role in distinguishing the critical points. The reducing values of the parameter $k$ means that the backreaction of the matter fields are negligible. A continuous slope is observed around the critical points which is  characteristic of the probe limit. In addition, exploring the nonlinear electrodynamic, the effects of the nonlinear parameter, $\beta$, is investigated. Finally the properties of conductivity are numerically explored in our model.
\end{abstract}

\section{Introduction}
To describe theories with strong coupling constants, one can use the AdS/CFT duality according to which, the physics of the conformal field theory (CFT) on the boundary of the AdS space time can be related to a gravitational theory in the bulk \cite{reff2, reff3, reff000}. One achievement of this duality is the establishment of the holographic entanglement entropy whereby one may calculate the entanglement entropy for a strongly coupled field theory using the Ryu-Takayanagi minimal surface \cite{reff5}-\cite{reffd6}. This method helps us understand the properties of various phase transitions in quantum field theories and in many-body physics \cite{reff000}-\cite{reff25}. A number of studies have investigated the entanglement entropy in various gravity theories \cite{reff7}-\cite{reff15}. Assuming a setup where an $AdS_{d+1}$  space is dual to a $CFT_{d}$ that lives on the boundary, the holographic entanglement entropy can be obtained by calculating the minimal area of the RT surface.
\begin{equation}\label{form1}
S=\frac{Area (\gamma_{RT})}{4G_{d+1}}.
\end{equation}
where, $G_{d+1}$ and $\gamma_{RT}$, respectively, are the Newton constant and the unique minimal surface anchored on the boundary of region $A$ in the $AdS_{d+1}$ space.

Quantum complexity was proposed as a tool for understanding the interior of black holes \cite{reff10550}. Complexity expresses how difficult it is to go from a given state to another. For an eternal black hole, the complexity is proportional to the spatial volume where the time slice connects two boundaries through the Einstein-Rosen bridge \cite{reff1045}.
Holographic complexity was defined in \cite{reff1035} as one that is holographically related to the volume enclosed by the minimal surface as follows:
\begin{equation}\label{form7007}
C=\frac{V(\gamma)}{8\pi l G_{d+1}}.
\end{equation}
where, $l$ is the radius of the curvature of the background spacetime and, in this work, it is the radius of the curvature of the $AdS$ spacetime.

On the other hand, it is well known that the AdS/CFT duality is an approach  for describing the physics of high $T_{c}$ superconductors.
Breaking $U(1)$ symmetry in the bulk theory we can construct a holographic superconductor by obtaining a scalar hair condensate  in the bulk theory at a temperature, $T$, below the critical one, $T_{c}$. For $T$ greater than $T_{c}$, there is no scalar hair which is called the normal phase.

The holographic entanglement entropy can be used to explore the metal/superconductor phase transition. Albash and Johnson found that the entanglement entropy in the metal phase is grater than the superconductor one \cite{reff27}. It can be seen that the entanglement entropy has a discontinuous slope at a second order of phase transition. In addition , when the first order of phase transition  occurs, one can see a jump in the entanglement entropy. Also, it has been suggested that holographic subregion complexity ($RT$ volume) can be used as a useful tool for probing superconductor phase transitions. Therefore, both holographic entanglement entropy and holographic subregion complexity are useful quantities to identify  superconductor phase transitions \cite{reff1025, reff1015, reff010, reff020}.

In \cite{reff1005}, the authors suggested a topological invariant of the volume enclosed by $RT$ surface, $\Sigma$. This quantity is defined as follows:
\begin{equation}\label{form505}
Invariant{C}[R]\varpropto \int _{\Sigma}R \sqrt{-g}d\sigma ,
\end{equation}
where, $R$ is the Ricci scalar. We call the above quantity as an invariant topological quantity. In the $AdS_{4}$ space, there are other topological invariants for this volume such as the following ones:
\begin{eqnarray}\label{form50105}
Invariant{C}[R_{\mu \nu}R^{\mu \nu}]&\varpropto & \int _{\Sigma}R_{\mu \nu}R^{\mu \nu} \sqrt{-g}d\sigma , \nonumber\\
Invariant{C}[R_{\mu \nu \rho \sigma}R^{\mu \nu \rho \sigma}]&\varpropto & \int _{\Sigma}R_{\mu \nu \rho \sigma}R^{\mu \nu \rho \sigma} \sqrt{-g}d\sigma .
\end{eqnarray}
where, $R_{\mu \nu}$ and $R_{\mu \nu \rho \sigma}$ are Ricci and Riemannian tensors, respectively.
In addition, motivated by the above consideration, we use the definition for holographic entanglement entropy and consider another topological invariant in the $AdS_{4}$ space for the $RT$ surface as follows:
\begin{equation}\label{form606}
Invariant{S}[R]\varpropto \int_{A} R\sqrt{-g}da.
\end{equation}
 In this work, we use the topological invariants related to the $RT$ surface and volume to identify phase transition points.
The interpretation of these topological invariants in the dual field theory are not yet given and remain to be a challenge. Plotting these invariant quantities as a function of temperature, a discontinuity slope is detected at critical points which specifies the second order of phase transition and shows the phase transition in the metal/superconductor system more clearly than the holographic entanglement entropy and holographic complexity do. Therefore, these topological invariants are useful tools for probing the superconductor phase transition. Another example of using the topological invariants of the $RT$ surface has been studied in \cite{reff0330, reff040}. In addition, we consider the Lifshitz black hole background that leads to the non-relativistic behaviour at finite temperatures for the boundary CFT. We also consider the nonlinear electrodynamic and study the effect of nonlinear parameter, $\beta$, on our results. The effect of matter fields becomes negligible as a result of decreasing values of the backreaction parameter, $k$. As we expect, phase transition is not so clearly distinguished; rather, we have a continuous slope around the critical point, indicating that we are in the probe limit. It will be even more difficult to observe the phase transition point if the Lifshitz parameter, $z$, increases. In addition, the non linear parameter, $\beta$, has a different effect on metal and superconductor phases such that, the invariant quantities increase in the superconductor phase with increasing $\beta$ while an inverse behaviour is observed in the metal phase. Finally, a numerical method is used to investigate the behaviour of conductivity and condensation operator in the general nonlinear electrodynamic.

This paper is organized as follows. In Section 2, we obtain the equations of motion and the critical temperature in our model. We investigate the superconductor phase transition from the topological invariant quantities of the RT surface and the volume enclosed by it in Section 3. The behaviour of the conductivity and condensation operators in our model are numerically studied in Section 4. We will summarize our results in the concluding section.

\section{Equations of motion and critical points in metal/super-\ conductor systems}\label{cap2}
The action in the Lifshitz background with a negative cosmological constant and for the non-linear electromagnetic field is \cite{reff28, reff29}:
\begin{equation}\label{form2}
I=\int {d^{4}x \sqrt{-g}\left[ \frac{1}{2 k^{2}}\left(R-2\Lambda -\frac{1}{2}\partial_{\mu}\Phi \partial^{\mu}\Phi-\frac{1}{4}e^{\lambda \Phi}F_{\mu\nu}F^{\mu\nu}\right) +L_{m} \right]},
\end{equation}
\begin{equation}\label{form3}
L_{m}=F+\beta^{2}F^{2}-\mid \nabla\psi -iqA\psi\mid^{2}-m^{2}\psi^{2},
\end{equation}
where, $F=-\frac{1}{4}F^{\mu\nu}F_{\mu\nu}$. $\Lambda$ is a cosmological constant, $\Lambda=-\frac{(z+1)(z+2)}{2l^2}$, and $k^{2}=8 \pi G_{4}$ where $G_{4}$ is the gravitational constant. $z$, $\beta$, and $A$ are the Liftshitz parameter, the nonlinear parameter, and the gauge field, respectively. The scalar field with a mass $m$ and a charge $q$ is represented by $\psi$. $\lambda$ can be related to $z$ by $\lambda^{2}=\frac{4}{z-1}$ and $\Phi $ is the scalar field to be obtained from the following equations of motion:
\begin{eqnarray}
\partial_{\mu}\left(\sqrt{-g}e^{\lambda \Phi}F^{\mu \nu}\right)=0, \nonumber\\
\partial_{\mu}\left(\sqrt{-g}\partial^{\mu} \Phi\right)-\frac{\lambda}{4}\sqrt{-g}e^{\lambda \Phi}F_{\mu \nu}F^{\mu \nu}=0.
\end{eqnarray}
In this work, the probe limit corresponds to $k^{2}\rightarrow 0$. In this limit, we have $\partial_{r}\Phi \partial_{r}\Phi=\frac{4(z-1)}{r^{2}}$ which can be obtained from the combination of $tt$ and $rr$ components of Einstein equations of motion. It should be noted that the  backreaction effect does not change this relation. The only non vanishing component of the field strength is $F_{rt}=\widehat{q}e^{-\lambda \Phi}r^{z-3}$, in which $\widehat{q}$ is a charge and can be related to $z$ as $\widehat{q}=2l^{2}(z+1)(z+2)$.

Varying the action with respect to the metric, Einstein equations of motion is given by
\begin{eqnarray}
R^{\mu \nu}&-&\frac{g^{\mu \nu}}{2}R-\frac{(z+1)(z+2)}{2l^2}g^{\mu \nu}-\frac{1}{2}\partial_{\mu}\Phi \partial_{\nu}\Phi \nonumber\\
&-&\frac{1}{2}e^{-\lambda \Phi}F_{\mu \rho}F_{\nu}^{\rho}+\frac{1}{8}g_{\mu \nu}e^{\lambda \Phi}F_{\mu \nu}F^{\mu \nu}=k^{2}T^{\mu \nu},
\end{eqnarray}
where, the energy momentum tensor is as follows:
\begin{eqnarray}
T^{\mu \nu}&=&g^{\mu \nu}\left(F+\beta^{2}F^{2}\right)+\left(1-\beta^{2}F^{2} \right)F_{\sigma}^{\mu}F^{\sigma \nu}-m^{2}g^{\mu \nu}|\psi |^{2}\nonumber\\
 &-&g^{\mu \nu}|\nabla \psi -iqA \psi |^{2}+\left[\left(\nabla^{\nu}-iqA^{\nu} \right) \psi^{*}\left( \nabla^{\mu}-iqA^{\mu}\right) \psi +\mu \leftrightarrow \nu\right].
\end{eqnarray}
Also, we have other equations of motion as in the following:
\begin{equation}
\nabla_{\mu}\left(F^{\mu \nu}-\beta^{2}F^{\mu \nu}F \right)=iq\left( \psi^{*}\left( \nabla^{\nu}-iqA^{\nu}\right) \psi-\psi\left(\nabla^{\nu}+iqA^{\nu} \right) \psi^{*}\right),
\end{equation}
\begin{equation}
\left( \nabla_{\mu}-iqA_{\mu} \right) \left( \nabla^{\mu}-iqA^{\mu}\right) \psi - m^{2}\psi=0.
\end{equation}

We choose both the $AdS$ radius, $l$, and the charge of the scalar field, $q$ , equal to 1. The  Liftshitz plane-symmetric black hole in the presence of the backreaction effect is given by:

\begin{equation}\label{form4}
ds^{2}=-r^{2z}f(r)e^{-\chi(r)}dt^{2}+\frac{dr^{2}}{r^{2}f(r)}+r^{2}(dx^{2}+dy^{2}).
\end{equation}
where, $f(r)$ is the metric function and $\chi(r)$ represents the back reaction effect. The Hawking temperature with the radius of event horizon, $r_{h}$, is
\begin{equation}\label{form410}
T=\frac{r_{h}^{z+1}f'(r_{h})e^{-\chi(r_{h})/2}}{4\pi}.
\end{equation}
We choose the following ansatz for the vector and scalar fields:
\begin{equation}\label{form5}
A_{\mu}=\left( \varphi(r),0,0,0\right), \  \  \  \psi=\psi(r).
\end{equation}
Varying action (\ref{form2}) with respect to the scalar field $\psi(r)$, gauge field $\varphi(r)$, and the above metric, the following equations of motion are obtained:

\begin{equation}\label{form6}
\psi''(r)+\left(\frac{f'(r)}{f(r)}+\frac{(z+3)}{r}-\frac{\chi'(r)}{2} \right) \psi'(r)+\left(\frac{e^{\chi(r)}\varphi^{2}(r)}{u^{2z}f(r)^{2}}-\frac{m^{2}}{f(r)}\right) \psi(r)=0,
\end{equation}
\begin{eqnarray}\label{form7}
&\varphi''(r)&\left(1+4\beta^{2}e^{\chi(r)}r^{2-2z}\varphi'^{2}(r)\right)+2\beta^{2}e^{\chi(r)}r^{2-2z}\chi'(r)\varphi'^{3}(r)+2\beta^{2}\left(2-2z\right)e^{\chi(r)}r^{1-2z}\varphi'^{3}(r)\nonumber\\
&+&\varphi'(r)\left(\frac{3-z}{r}+\frac{\chi'(r)}{2} \right) -2\psi^{2}(r)\varphi(r)\left(1-2\beta^{2}e^{\chi(r)r^{2-2z}} \varphi'^{2}(r)\right)=0,
\end{eqnarray}
\begin{eqnarray}\label{form9}
2rf'(r)&-&rf(r)\chi'(r)+\left( 2z+4\right) f(r)-z^{2}\left( 1-e^{\chi(r)}\right) +2(1+e^{\chi(r)})+z\left( 3-e^{\chi(r)}\right) \nonumber\\
&+&2k^{2}\left( m^{2}\psi^{2}(r)+\frac{1}{2}e^{\chi(r)}r^{2-2z}\varphi'^{2}(r)+\beta^{2}e^{2\chi(r)}\varphi'^{4}(r)r^{4-4z}\right)=0,
\end{eqnarray}
Also, combination of $tt$ and $rr$ components of the Einstein equations leads to
\begin{equation}\label{form8}
\chi'(r)+2k^{2}r\left(\frac{\psi^{2}(r)\varphi^{2}(r)e^{\chi(r)}}{f^{2}(r)r^{2z+2}}+\psi'^{2}(r) \right) =0.
\end{equation}

We can use a transformation such as $u=\frac{r_{h}}{r}$ to turn the equations of motion into the following forms:

\begin{eqnarray}\label{form610}
&u^{2}&f(u)\psi''(u)-u(z+1)f(u)\psi'(u)-m^{2}\psi(u)+u^{2}f'(u)\psi'(u)\nonumber\\
&-&\frac{1}{2}u^{2}f(u)\chi'(u)\psi'(u)+\frac{e^{\chi(u)}\psi(u)\varphi^{2}(u)u^{2z}}{f(u)rh^{2z}}=0,
\end{eqnarray}
\begin{eqnarray}\label{form710}
&u^{2}&f(u)\varphi''(u)+\frac{1}{2}u^{2}f(u)\chi'(u)\varphi'(u)+u(z-3)f(u)\varphi'(u)\nonumber\\
&+&\frac{2\beta^{2}u^{4+2z}f(u)e^{\chi(u)}}{r_{h}^{2z}}\chi'(u)\varphi'^{3}(u)-2\psi^{2}(u)\varphi(u)
\left( 1-\frac{2\beta^{2}u^{2+2z}e^{\chi(u)}\varphi'^{2}(u)}{r_{h}^{2z}}\right)\nonumber\\
&+&\frac{4\beta^{2}u^{4+2z}f(u)}{r_{h}^{2z}}e^{\chi(u)}\varphi'^{2}(u)\varphi''(u)+\frac{4\beta^{2}u^{3+2z}(z-1)f(u)}{r_{h}^{2z}}e^{\chi(u)}\varphi'^{3}(u)=0,
\end{eqnarray}
\begin{eqnarray}\label{form910}
&2&u^{2}f'(u)+u^{2}f(u)\chi'(u)-u(2z+1)f(u)-3uf(u)+uz^{2}(1-e^{\chi(u)})\nonumber\\
&-&3zu-zue^{\chi(u)}+\frac{k^{2}u^{3+2z}e^{\chi(u)}\varphi'^{2}(u)}{r_{h}^{2z}}-\frac{2\beta^{2}k^{2}u^{5+4z}e^{2\chi(u)}\varphi'^{4}(u)}{r_{h}^{4z}}\nonumber\\
&-&\frac{2k^{2}u^{3+2z}e^{\chi(u)}\varphi'^{2}(u)}{r_{h}^{2z}}-2k^{2}m^{2}u\psi^{2}(u)-2u(e^{\chi(u)}+1)=0,
\end{eqnarray}
\begin{equation}\label{form810}
\chi'(u)+\frac{2k^{2}}{f^{2}(u)}\left(\psi^{2}(u)\varphi^{2}(u)e^{\chi(u)}\frac{u^{2z-1}}{r_{h}^{2z}}+\psi'^{2}(u)uf^{2}(u) \right) =0.
\end{equation}

To obtain temperature in the metal phase the scalar field $\psi$ should be turned off and $\chi$ is found as a constant from equation (\ref{form810}). Therefore, by taking the gauge field to be $\varphi=\mu - \rho (\frac{u}{rh})^{2-z}$ near the boundary and considering  the regularity condition in equation (\ref{form710}) for the gauge field near the event horizon, $\varphi(1)=0$, we can obtain the metric function from (\ref{form910}) as follows:
\begin{equation}\label{form222}
f(u)=-\frac{1}{8}u^{z+2}\left(8(1-u^{-z-2})+\frac{4k^{2}\rho^{2}(z-2)^{2}}{(6-5z)r_{h}^{4}}(1-u^{6-5z})+\frac{\beta^{2}k^{2}\rho^{4}(2-z)^{4}}{(-14+9z)r_{h}^{8}}(1-u^{14-9z}) \right),
\end{equation}
where, $\rho$ is the charge density and $\mu$ is the chemical potential in the boundary. As can be seen, at the event horizon, we have $f(1)=0$. Therefore, the temperature in the metal phase can be obtained from Eqs. (\ref{form410}) and (\ref{form222}) as bellow
\begin{equation}\label{form1022}
T=\frac{r_{h}^{z}}{4\pi}\left((z+2)-\frac{k^{2}\rho^{2}(z-2)^{2}}{2r_{h}^{4}}+\frac{\beta^{4}k^{2}\rho^{4}(z-2)^{4}}{8r_{h}^{4}} \right).
\end{equation}

In the superconductor phase, the scalar field, $\psi$, is not equal to zero. Near the boundary, where $u \rightarrow 0$, the scalar field behaves as $\psi \approx \frac{\psi_{-}u^{\Delta_{-}}}{r_{h}^{\Delta_{-}}}+\frac{\psi_{+}u^{\Delta_{+}}}{r_{h}^{\Delta_{+}}}$. Here, we assume that $\psi_{+}$ is the source of the dual operator $<O>$ and the condensation operator of the dual field theory is, therefore, defined as $<O>=\psi_{-}$. The regularity condition in equation (\ref{form6}) for the scalar field is $\psi(r_{rh})=\frac{f'(r_{rh})\psi'(r_{rh})}{m^{2}}$.

In order to obtain temperature in the superconductor phase, we need to solve Eq. (\ref{form910}) perturbatively. At first, we expand the gauge field $\varphi(u)$, scalar field $\psi(u)$, the metric function $f(u)$ and $\chi(u)$ near the critical point to the following forms \cite{reff00}:
\begin{equation}\label{form13}
\varphi(u)=\varphi_{0}(u)+\varepsilon^{2}\varphi_{2}(u)+\varepsilon^{4}\varphi_{4}(u)+...,
\end{equation}
\begin{equation}\label{form14}
\psi(u)=\varepsilon \psi_{1}(u)+\varepsilon^{3}\psi_{3}(u)+\varepsilon^{5}_{5}(u)+...,
\end{equation}
\begin{equation}\label{form15}
f(u)=f_{0}(u)+\varepsilon^{2} f_{0}^{2}(u)+\varepsilon^{4}f_{0}^{4}(u)+...,
\end{equation}
\begin{equation}\label{form16}
\chi(u)=\varepsilon^{2}\chi_{2}(u)+\varepsilon^{4}\chi_{4}(u)+... .
\end{equation}
The expansion parameter, $\varepsilon$, can be defined as $\varepsilon \equiv <O>$, which is a small value near the critical point. Substituting the above equations into the equations of motion, we obtain the following relation for the gauge field and the metric function obtained from Eqs. (\ref{form710}) and  (\ref{form910}) at zero order of the expansions of the fields and up to the second order of $\beta$,

\begin{equation}\label{form17}
u\varphi_{0}''(u)+u(z-1)\varphi_{0}'(u)+4\beta^{2}u^{3}(\frac{r_{h}}{u})^{-2z}\left(u\varphi_{0}'^{2}(u)\varphi_{0}"(u)+(z+1)\varphi_{0}'^{3}(u)\right) =0,
\end{equation}
\begin{eqnarray}\label{form190}
f_{0}(u)&=&2u(z+2)-2u^{2}f'_{0}(u)+3\beta^{2}k^{2}u^{5}(\frac{r_{h}}{u})^{-4z}\varphi_{0}^{4}(u)\nonumber\\
&+&k^{2}u^{3}(\frac{r_{h}}{u})^{-2z}\varphi_{0}'^{2}(u)-2u(z+2).
\end{eqnarray}
To obtain $\varphi_{0}(u)$ from the above equation, $\varphi_{0}'(u)$ is replaced by $y(u)$  in Eq. (\ref{form17}) and the resulting equation is integrated in the interval [0,1]. Finally, the following equation is obtained:

\begin{equation}\label{form199}
\varphi_{0}(u)=\frac{u^{2-z}-1}{2-z}C-\frac{2\beta^{2}(u^{6-z}-1)}{(6-z)r_{h}^{2z}}C^{3}.
\end{equation}
where, $C$ is the integration constant. Replacing (\ref{form199}) into (\ref{form190}) and keeping the second order of $\beta$, we  obtain the following relation for $f_{0}(u)$:

\begin{equation}\label{form010}
f_{0}(u)=-\frac{1}{8}u^{z+2}\left(8(1-8u^{-2-z})+\frac{4k^{2}C^{2}(z-2)^{2}}{6-5z}(1-u^{6-5z})+\frac{\beta^{2}k^{2}C^{4}(z-2)^{4}}{-14+9z}(1-u^{14-9z}) \right).
\end{equation}
On the other hand, in the first order approximation, the scalar field near the boundary, $u\rightarrow 0$, takes the following form:

\begin{equation}
\psi_{1}\approx  \frac{\psi_{-}u^{\Delta_{-}}}{r_{h}^{\Delta_{-}}}+\frac{\psi_{+}u^{\Delta_{+}}}{r_{h}^{\Delta_{+}}}.
\end{equation}

Matching the behaviour of $\psi_{1}$ near the boundary, one can define the following relation for $\psi_{1}(u)$

\begin{equation}\label{form188}
\psi_{1}(u)=\frac{\langle O \rangle}{\sqrt{2}r_{h}^{\Delta}}u^{\Delta}F(u),
\end{equation}

\noindent where, $F(u)$ is a trial function, $F(u)=1-\alpha u^2$, that satisfies the boundary condition near the boundary of the $AdS$ as $F(0)=1$ and $F'(0)=1$ \cite{reff11111}. Also, $\langle O \rangle$ can be interpreted as the condensation operator in the superconductor phase. Substituting (\ref{form188}) into Eq. (\ref{form610}) yields:

\begin{equation}\label{form26}
NF''+N'F'+PF+C^{2}QF=0,
\end{equation}
where,
\begin{equation}\label{form27}
N=u^{2\Delta -z-1}f_{0}(u),
\end{equation}
\begin{equation}\label{form28}
P=\frac{\left(-m^{2}+\Delta (-2-z-\Delta)\right)f_{0}(u)+u\Delta f_{0}'(u)}{u^{2}}u^{2\Delta -z-1},
\end{equation}
\begin{equation}\label{form29}
Q=\frac{(1-n^{2-z})^{2}}{f_{0}(u)}u^{-1-z+2\Delta}-\frac{C^{2}\beta^{2}u^{2\Delta -3-z}(z-2)(u^{z}-u^{2})(u^{z}-u^{6})}{(z-6)f_{0}(u)}.
\end{equation}
According to the Sturm-Liouville eigenvalue problem, the eigenvalue $C^{2}$ can be determined by minimizing the expression:
\begin{equation}\label{form30}
C ^{2}=\frac{\int {N(F'^{2}-PF^{2})du}}{\int{NQF^{2}du}}.
\end{equation}

The value for $C$ can be determined numerically using the iterative method. To simplify the calculations, we introduce the backreaction parameter as follows \cite{reff12221}:
\begin{equation}
k=k_{n}=n \Delta k \ \ , \ \ \ n=0,1,...n_{max}  \ \ and \ \ \  \Delta k=k_{n+1}-k_{n}.
\end{equation}
where, $\Delta k$ is the size of the iterative method. We obtain $C^2$ up to the second order of $\beta^2$. Using the iterative procedure, we also define the following expressions:
\begin{eqnarray}
k^{2}C^{2}&=&k_{n}^{2}C^{2}=k_{n}^{2}C_{k_{n-1}}^{2}+O\left(\Delta \kappa^4 \right),\nonumber\\
\beta^{2}C^{2}&=&\beta^{2}C_{\beta^{2}=0}^{2}+O\left( \beta^4\right),\nonumber\\
\beta^{2}k^{2}C^{4}&=&\beta^{2}k_{n}^{2}C_{k_{n-1},\beta^{2}=0}^{4}+O\left( \beta^4\right)+O\left(\Delta \kappa^4 \right).
\end{eqnarray}

The first step starts when $n = 0$ or the backreaction parameter is zero; i.e., $k_{-1} = 0$, $C_{k_{-1}}=0$, and $C_{k_{-1},\beta^{2}}=0$. In this step, we are in the probe limit. In the next step, $n=1$, the back reaction parameter has a small value $k_{1}=\Delta k $ and the value of $C_{k_{0}}$ and $C_{k_{0},\beta^2=0}$ can be replaced with those obtained from the previous step. This procedure continues until the temperature in the superconductor phase is less than that at the critical point. Thus, to obtain the temperature in the superconductor phase, the numerically obtained value of $C$ should be replaced in relation (\ref{form010}), and (\ref{form410}) is used to obtain the temperature. This is captured by (\ref{form900}) below:

\begin{equation}\label{form900}
T=\frac{r_{h}^{z}}{4\pi}\left((z+2)-\frac{C^{2}k^{2}(z-2)^{2}}{2}+\frac{C^{4}\beta^{2}k^{2}(z-2)^{2}}{8} \right).
\end{equation}

The critical temperature is obtained based on the value of $C$ at the critical point. Thus, the definition for the gauge field near the boundary and at the critical point, $\varphi=\mu - \rho (\frac{u}{rh})^{2-z}$, is equated with (\ref{form199}) at the boundary ($u = 0$), to obtain an expression for $C$:
\begin{equation}
C=\frac{\rho}{r_{hc}^{2}}.
\end{equation}
Replacing the above relation into  (\ref{form010}) and using  (\ref{form410}) yield the critical temperature as follows:

\begin{equation}\label{form10002}
T_{c}=\frac{\sqrt{(\frac{\rho}{C})^{z}}}{4\pi}\left((z+2)-\frac{C^{2}k^{2}(z-2)^{2}}{2}+\frac{C^{4}\beta^{2}k^{2}(z-2)^{2}}{8} \right).
\end{equation}

The holographic entanglement entropy and the holographic complexity have already been used for investigating the metal/superconductor phase transition. In the next section, we obtain these quantities for our model and see the behaviour of the superconductor phase transition. Also, we introduce other topological  invariants of the $RT$ surface and the volume enclosed by it and see how the critical point in the metal/superconductor system can be identified by these quantities.

\section{Topological invariants and the critical points}

\begin{figure*}
\centering
\begin{tabular}{ccc}
\rotatebox{0}{
\includegraphics[width=0.50\textwidth,height=0.18\textheight]{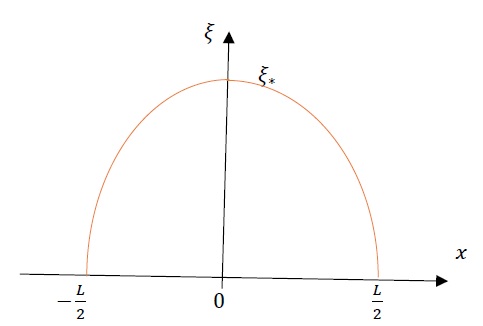}}\\
\end{tabular}
\caption{\label{fig111}minimal surface of a strip entangling surface of width $L$ }
\end{figure*}

Initially, the holographic entanglement entropy and the holographic complexity are obtained for a strip in our model \cite{reff7}-\cite{reff1035}, \cite{reff1025, reff1015}. In order to use the Poincare coordinates we choose a time slice in the metric (\ref{form4}) and replace the coordinate $r$ by $\frac{1}{\xi}$. Therefore, we have:

\begin{equation}
ds^{2}=\frac{d\xi^{2}}{\xi^{2}f(\xi)}+\frac{1}{\xi^{2}}(dx^{2}+dy^{2}).
\end{equation}
The strip is located at $\xi=0$ on the boundary. For a finite strip geometry along the $x$ direction, the subsystem $A$ can be described as $-\frac{L}{2}\leqslant x \leqslant\frac{L}{2}$ and $-\infty\leqslant y \leqslant\infty$, where $L$ is the size of region $A$. The minimal surface can be drawn as in Fig.(\ref{fig111}). Therefore, assuming $\xi$ as a function of $x$, we may obtain the induced metric on the minimal surface as follows:
\begin{equation}\label{form10}
ds^{2}_{\gamma_{RT}}=\frac{1}{\xi^{2}}\left(\left( (1+(\frac{\partial \xi}{\partial x})^{2}\frac{1}{f(\xi)}\right)dx^{2}+dy^{2}\right).
\end{equation}
The world volume of the above metric is the area of the minimal surface $\gamma_{RT}$ as in the following:
\begin{equation}\label{form1901}
A(\gamma_{RT})=Vol(\Re)\int_{-L/2}^{L/2}{\frac{dx}{\xi^{2}}\sqrt{1+(\frac{\partial \xi}{\partial x})^{2}\frac{1}{f(\xi)}}},
\end{equation}
And, the volume enclosed by the minimal surface $\gamma_{RT}$ is as follows:
\begin{equation}\label{form707}
V(\gamma_{RT})=Vol(\Re)\int_{\xi_{*}}^{\xi} \frac{x(\xi) d\xi}{\xi^{3}\sqrt{f(\xi)}}.
\end{equation}

To effect the minimal condition for the above surface, it is easy to regard $x$ as a time and the integrand of (\ref{form1901}) as a Lagrangian. Thus, the Hamiltonian can be written as follows:
 \begin{equation}\label{form1019}
 H=-\frac{1}{\xi^{2}\sqrt{1+\frac{\xi'^{2}}{f(\xi)}}},  \   \  \frac{dH}{dx}=0.
 \end{equation}
We consider $\xi_{*}$ as a maximum value of $\xi(x)$, where $\xi'(x)=0$ (Fig.\ref{fig111}). From the symmetry at $x=0$, we have $\xi_{*}=\xi(0)$. Examination of Eq. (\ref{form1019}) at $x=0$ yields the constant $H=-\frac{1}{\xi_{*}^{2}}$. Therefore, Eq. (\ref{form1019}) may be rewritten as (\ref{form102}) below:
\begin{equation}\label{form102}
\frac{d\xi}{dx}=-\frac{\sqrt{f(\xi)(\xi_{*}^{4}-\xi^{4})}}{\xi^{2}},
\end{equation}
or
\begin{equation}\label{form5555}
x(\xi)=-\int_{\xi_{*}}^{\xi}\frac{\xi^{2}d\xi}{\sqrt{f(\xi)(\xi_{*}^{4}-\xi^{4})}}.
\end{equation}
Integrating $dx$ from $0$ to $\frac{L}{2}$ in Eq. (\ref{form102}), we find a relation for the width $L$ as follows:
\begin{equation}\label{form1102}
\frac{L}{2}=\int_{\varepsilon}^{\xi_{*}}\frac{\xi^{2}d\xi}{\sqrt{f(\xi)(\xi_{*}^{4}-\xi^{4})}}.
\end{equation}
 where the UV cut off, $\varepsilon$, regularized the divergence coming from $\xi=0$. To obtain the volume enclosed by the $RT$ surface in Eq. (\ref{form707}), we need to obtain $\xi_{*}$ numerically from the above equation by fixing $L$. Then, $\xi(x)$ can be numerically calculated using Eq. (\ref{form5555}) to be used in Eq. (\ref{form707}) to determine volume numerically.

Substituting (\ref{form102}) into (\ref{form1901}) and replacing $\xi$ with $\frac{u}{r_{h}}$ transforms the area functional into (\ref{form1002}) below:
\begin{equation}\label{form1002}
A(\gamma_{RT})=2Vol(\Re)\int_{\varepsilon /r_{h}}^{u_{*}/r_{h}}{\frac{r_{h}^{2}u_{*}^{2}du}{u^{2}\sqrt{(u_{*}^{4}-u^{4})f(u)}}}.
\end{equation}
Therefore, (\ref{form1002}) may be used to obtain the entanglement entropy as follows:
\begin{equation}\label{form1000}
S=\frac{Vol(\Re)}{2G_{4}}\int_{\varepsilon /r_{h}}^{u_{*}/r_{h}}\frac{r_{h}^{2}u_{*}^{2}du}{u^{2}\sqrt{(u_{*}^{4}-u^{4})f(u)}}.
\end{equation}

\begin{figure*}
\centering
\begin{tabular}{ccc}
\rotatebox{0}{
\includegraphics[width=0.33\textwidth,height=0.16\textheight]{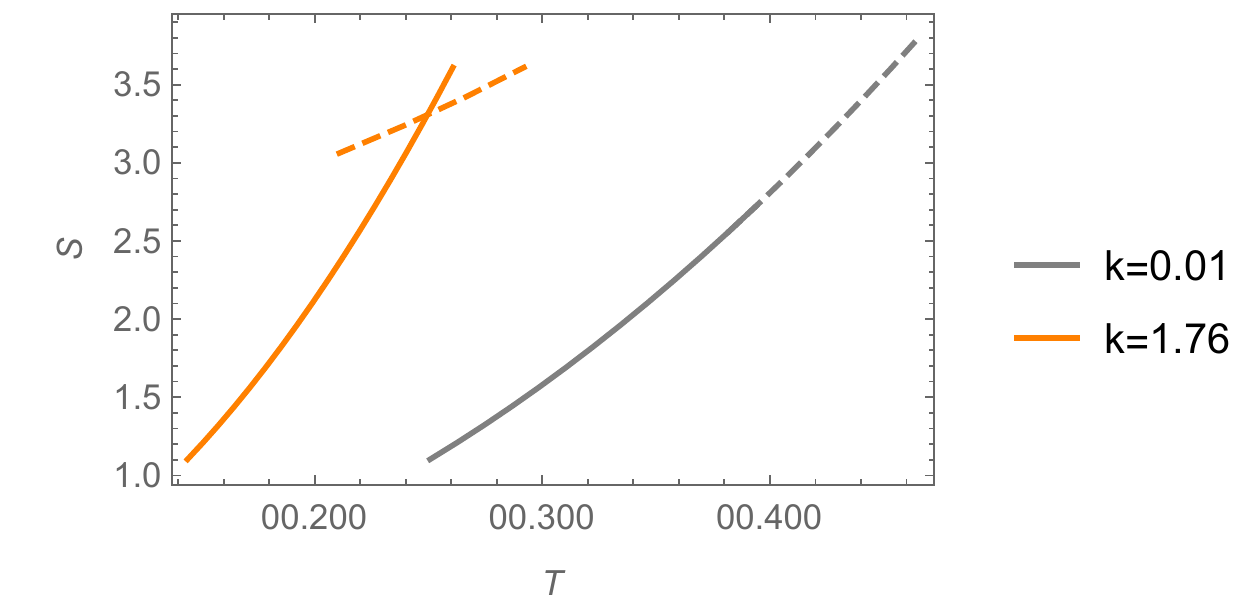}}&
\rotatebox{0}{
\includegraphics[width=0.33\textwidth,height=0.16\textheight]{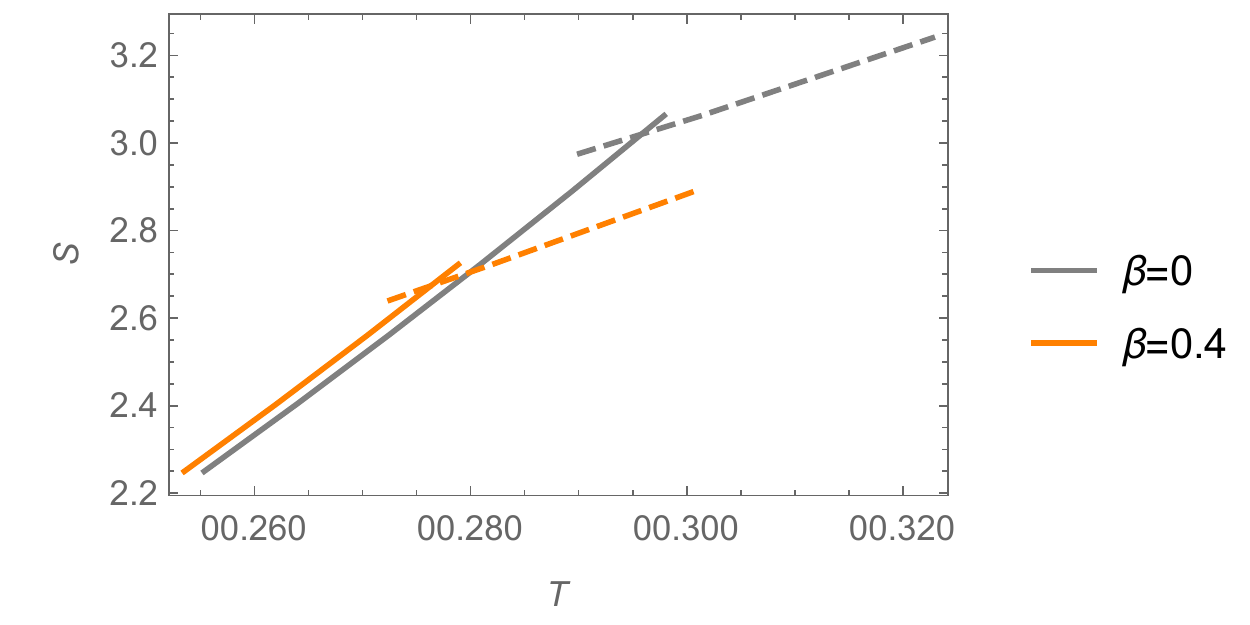}}&
\rotatebox{0}{
\includegraphics[width=0.33\textwidth,height=0.16\textheight]{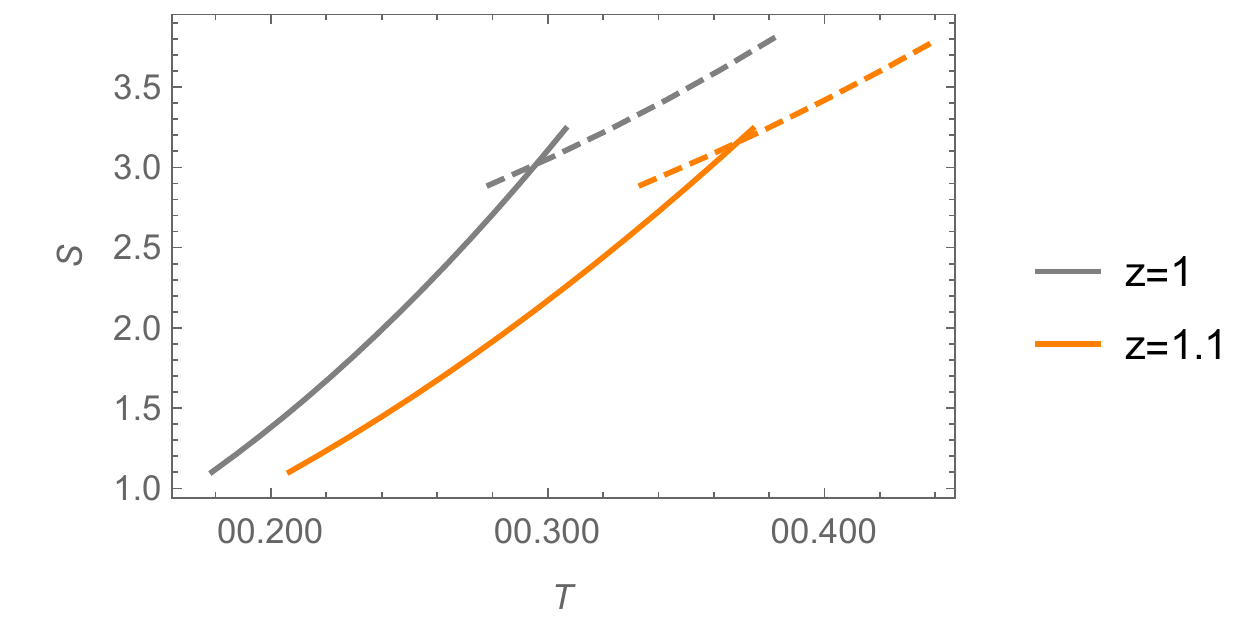}}\\
\rotatebox{0}{
\includegraphics[width=0.33\textwidth,height=0.16\textheight]{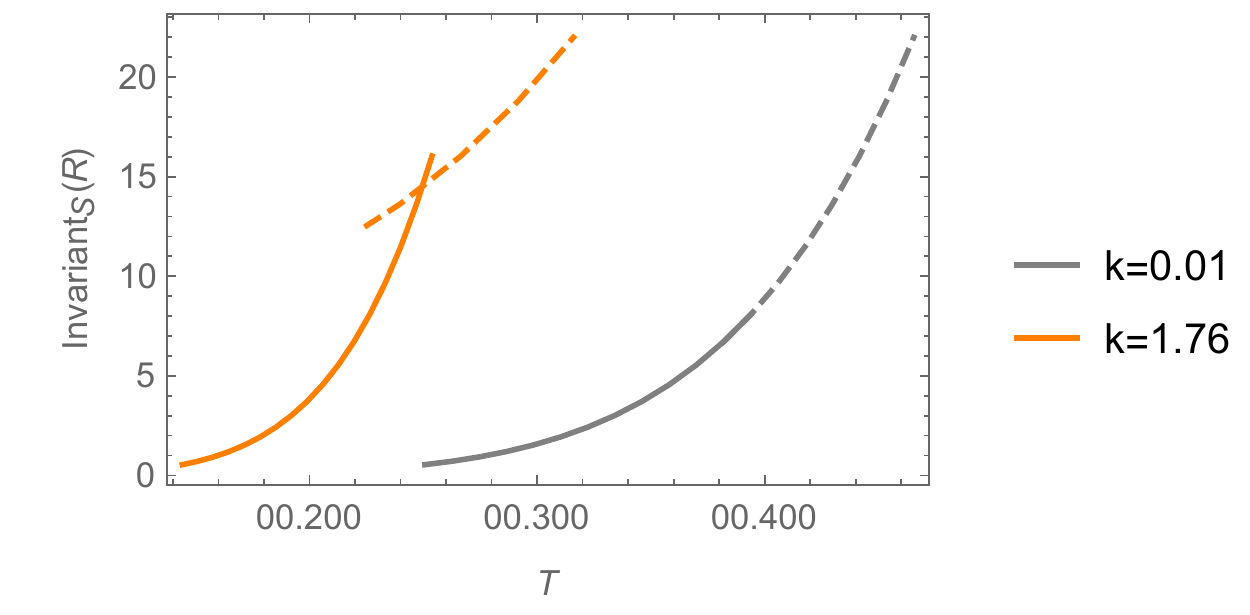}}&
\rotatebox{0}{
\includegraphics[width=0.33\textwidth,height=0.16\textheight]{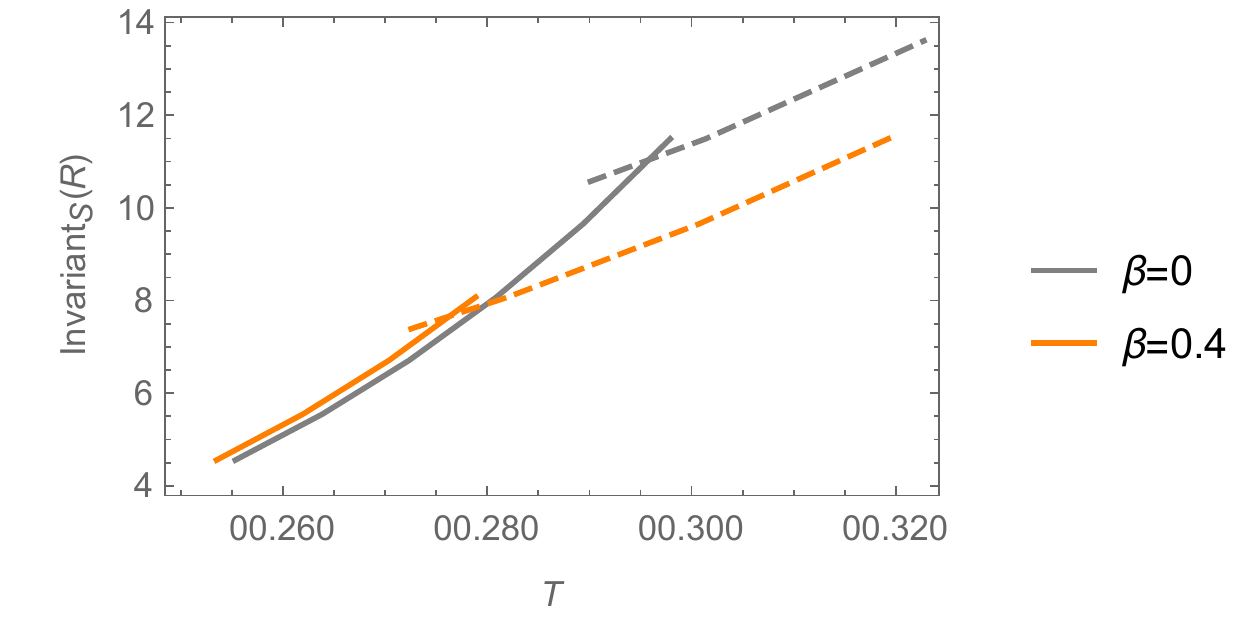}}&
\rotatebox{0}{
\includegraphics[width=0.33\textwidth,height=0.16\textheight]{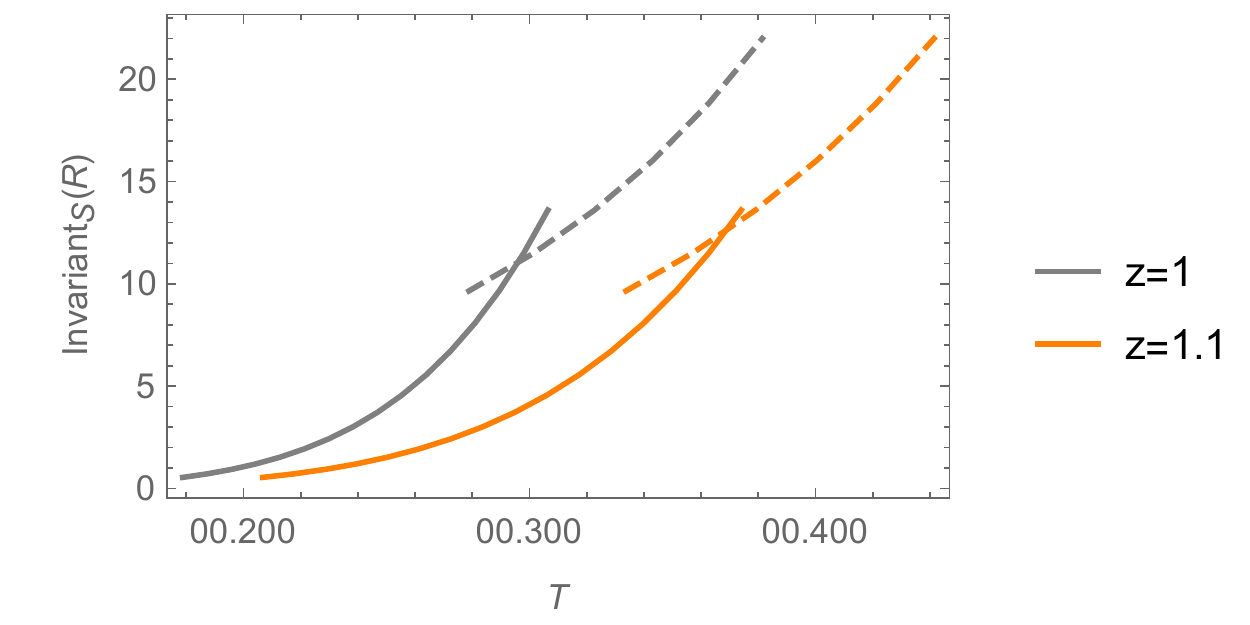}}\\
\end{tabular}
\caption{\label{fig1}The entanglement entropy at the top and a topological invariant quantity for the $RT$ surface at the bottom are shown as a functions of the temperature for different values of $k, \beta=0, z=1$ (left), $\beta, k=1.32, z=1$ (middle) and $z, \beta=0, k=1.32$ (right). The continuous lines represent the presence of the superconductor phase and the dashed lines represent the normal phase. In all these plots, $m^2$ and $\rho$ are chosen to be equal (-2) and (3), respectively.}
\end{figure*}

\begin{figure*}
\centering
\begin{tabular}{ccc}
\rotatebox{0}{
\includegraphics[width=0.33\textwidth,height=0.16\textheight]{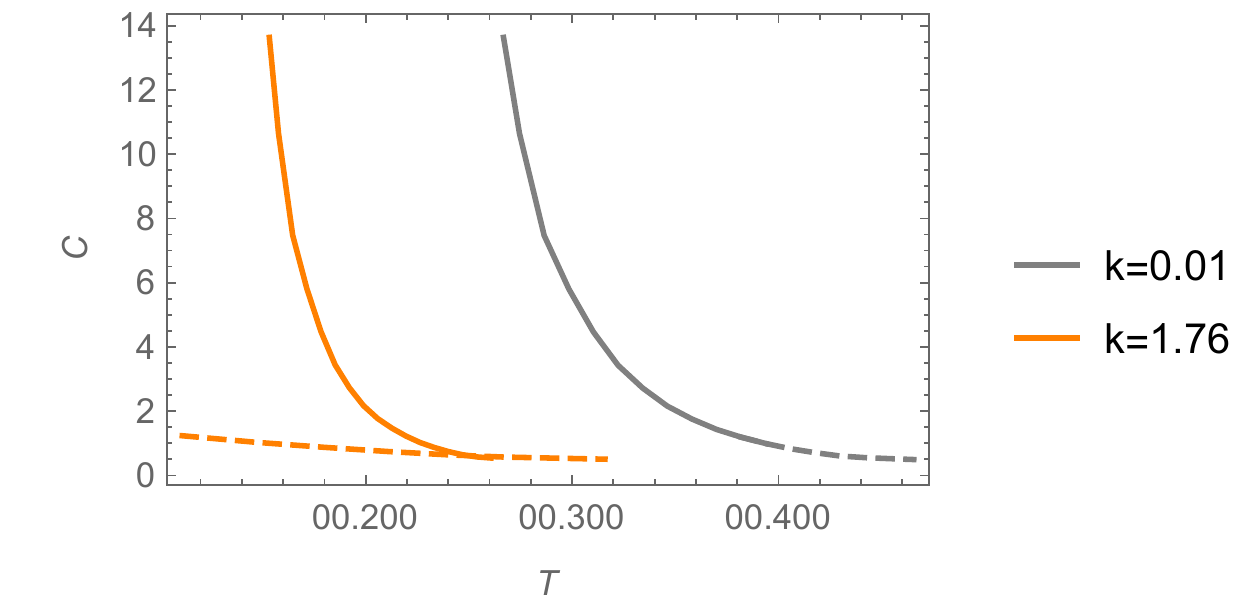}}&
\rotatebox{0}{
\includegraphics[width=0.33\textwidth,height=0.16\textheight]{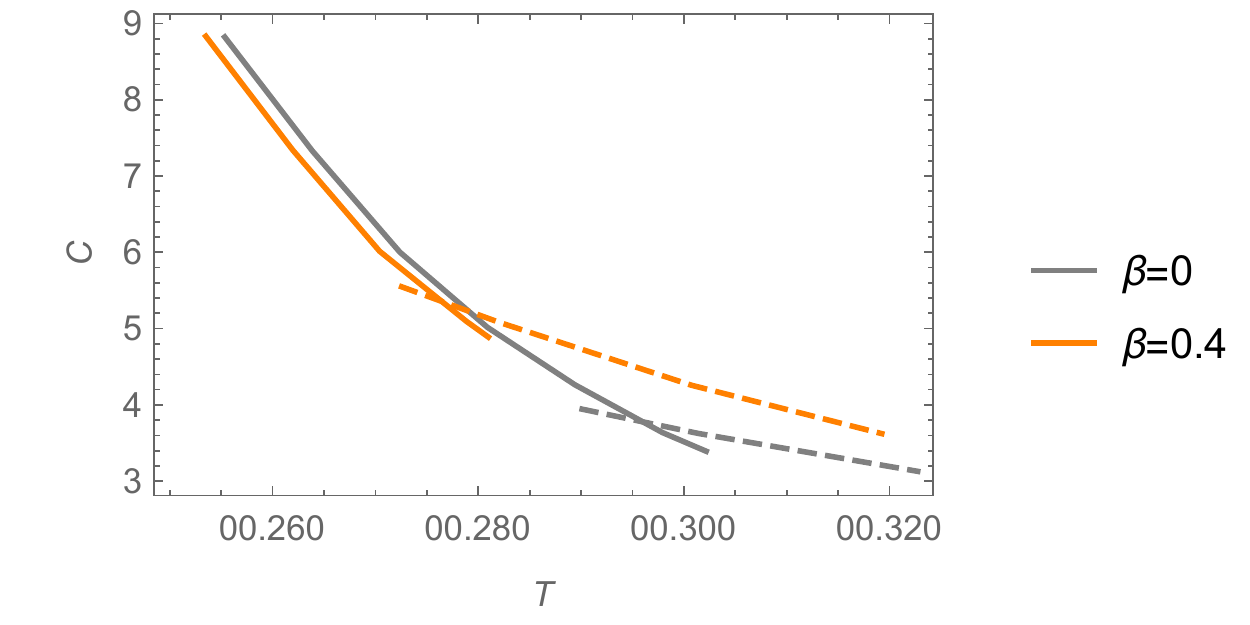}}&
\rotatebox{0}{
\includegraphics[width=0.33\textwidth,height=0.16\textheight]{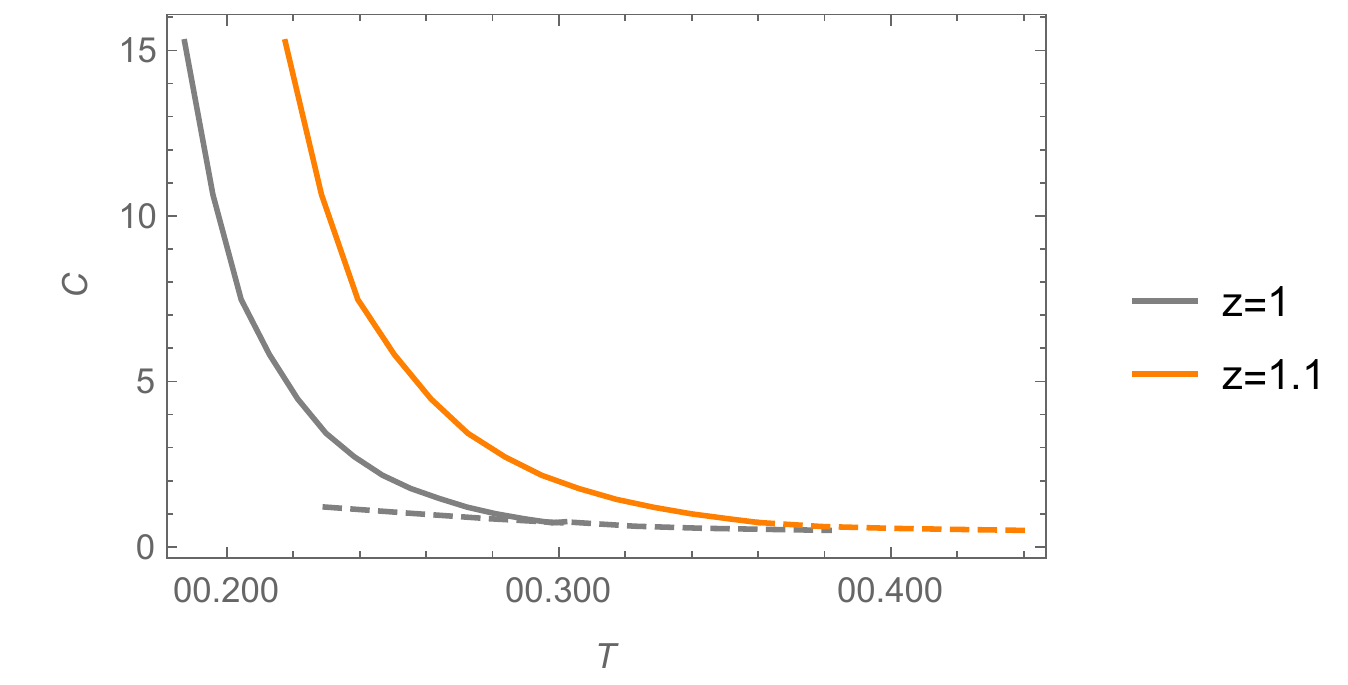}}\\
\rotatebox{0}{
\includegraphics[width=0.33\textwidth,height=0.16\textheight]{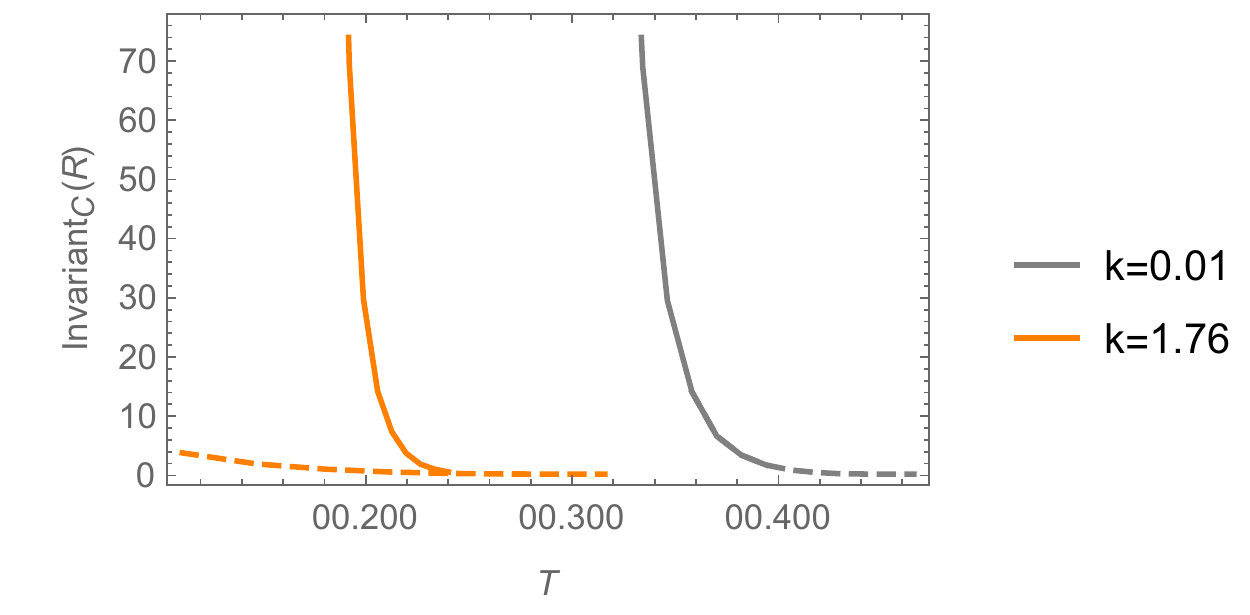}}&
\rotatebox{0}{
\includegraphics[width=0.33\textwidth,height=0.16\textheight]{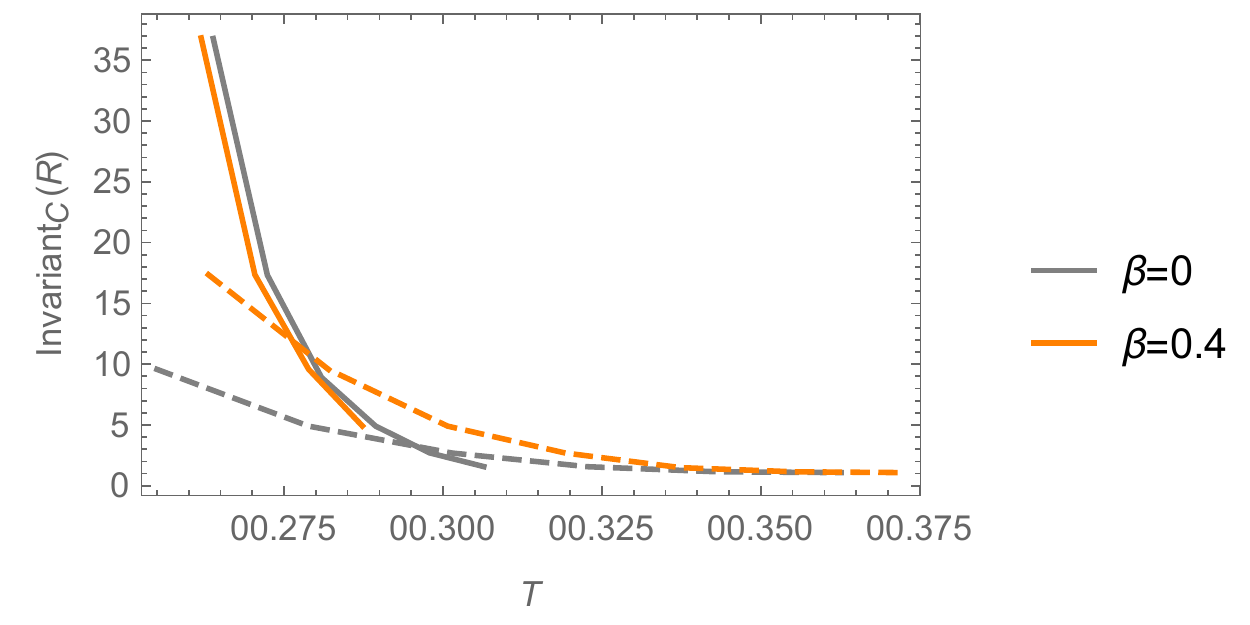}}&
\rotatebox{0}{
\includegraphics[width=0.33\textwidth,height=0.16\textheight]{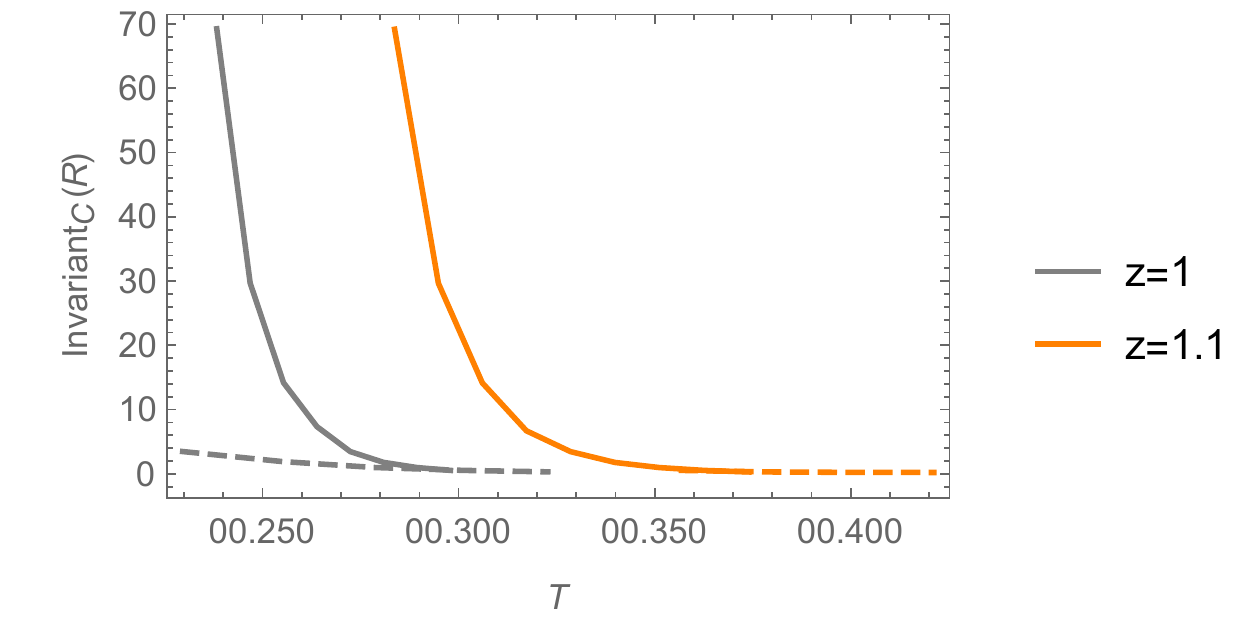}}\\
\rotatebox{0}{
\includegraphics[width=0.33\textwidth,height=0.16\textheight]{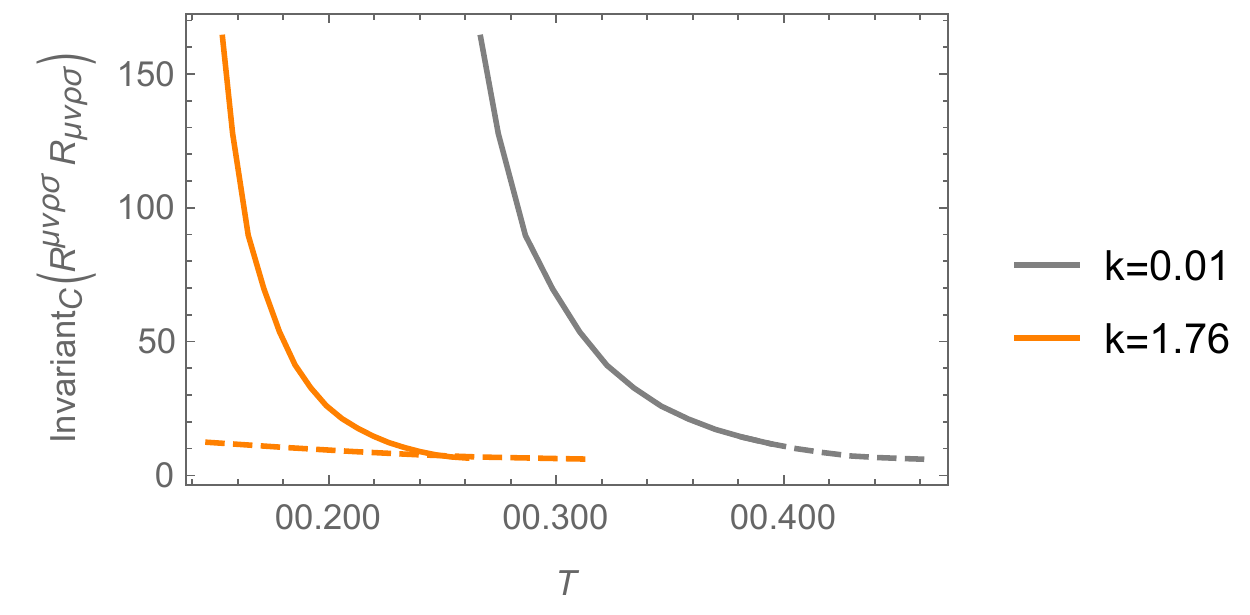}}&
\rotatebox{0}{
\includegraphics[width=0.33\textwidth,height=0.16\textheight]{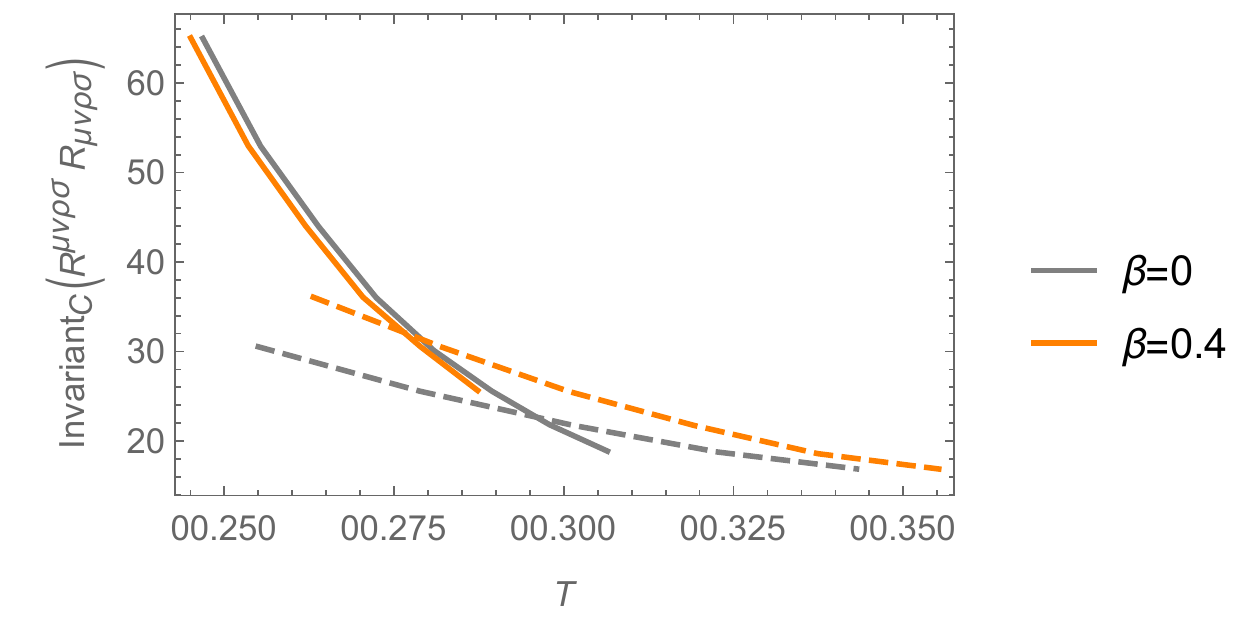}}&
\rotatebox{0}{
\includegraphics[width=0.33\textwidth,height=0.16\textheight]{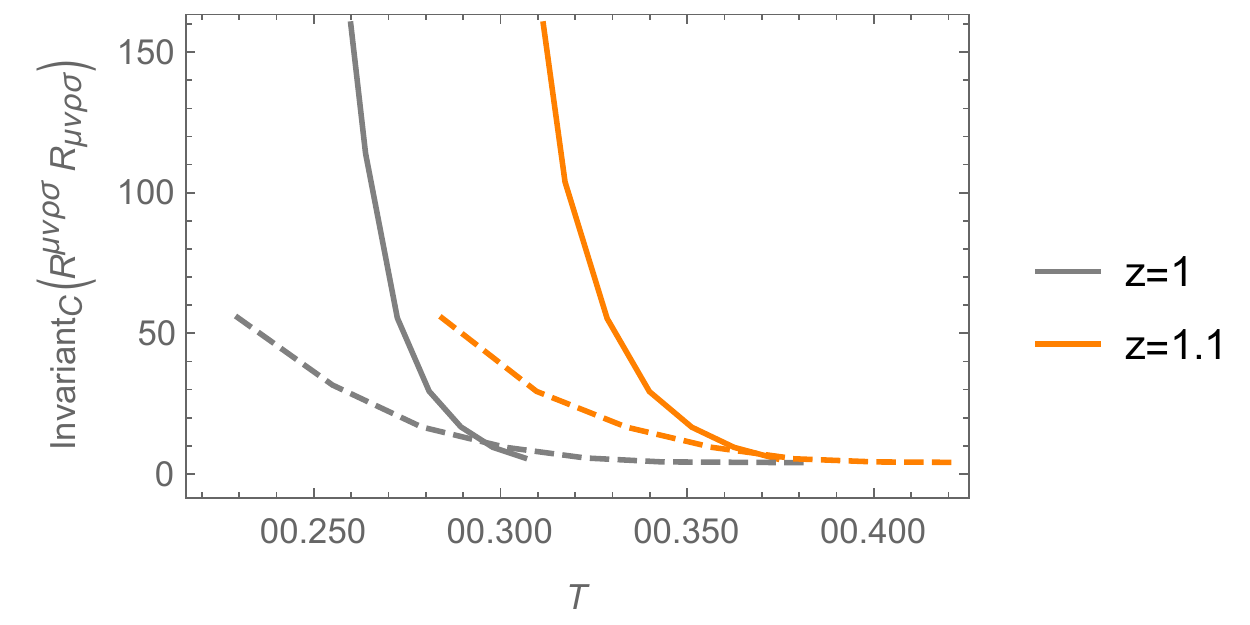}}\\
\end{tabular}
\caption{\label{fig22}The plots for complexity at the top and  topological invariant quantities for $R$ and $R_{\mu \nu \rho \sigma}R^{\mu \nu \rho \sigma}$ in the following rows, respectively, are presented as  functions of  temperature for different values of $k, \beta=0, z=1$ (left), $\beta, k=1.32, z=1$ (middle) and $z, \beta=0, k=1.32$ (right). The continuous lines represent the superconductor phase and the dashed ones represent the normal phase. In all these plots, $m^2$ and $\rho$ are chosen to be equal (-2) and (3), respectively.}
\end{figure*}

Replacing $\xi$ by $\frac{u}{r_{h}}$ in Relation (\ref{form707}), we obtain the subregion complexity as follows:
 \begin{equation}\label{form1100}
  C=\frac{Vol(\Re)}{8 \pi G_{4}l}\int_{u_{*}/r_{h}}^{\varepsilon /r_{h}}\frac{x(u)du}{u^{3}\sqrt{f(u)}},
 \end{equation}
 where,
 \begin{equation}
 x(u)=-\int_{u_{*}/r_{h}}^{\varepsilon /r_{h}}\frac{u^{2}du}{\sqrt{(u_{*}^{4}-u^4)f(u)}}.
 \end{equation}
The metric function $f(u)$ from Eq. (\ref{form222}) may be used for the metal phase and Eq. (\ref{form010}) for the superconductor phase. Also, the temperatures for both phases may be obtained from (\ref{form410}) and the critical temperature from (\ref{form10002}). The holographic entanglement entropy and holographic complexity are depicted in terms of temperature in the metal/superconductor system in Figs .(\ref{fig1}) and (\ref{fig22}).

We are now in a position to calculate some other topological invariants of the $RT$ surface and the volume enclosed by it. These topological invariants can be used to identify superconductor phase transition points more easily. Using the definition in Eq. (\ref{form606}) and considering the Relation (\ref{form1000}), we have the following expression for the invariant topological quantity of the $RT$ surface:

\begin{equation}\label{form333}
Invariant{S}[R]= \frac{Vol(\Re)}{4G_{4}}\int_{\varepsilon /r_{h}}^{u_{*}/r_{h}} R\frac{r_{h}^{2}u_{*}^{2}du}{u^{2}\sqrt{u_{*}^{4}-u^{4}f(u)}},
\end{equation}
where the Ricci scalar $R$ for the two dimensional $RT$ surface is as follows:
\begin{equation}
 R=\frac{r_{h}\left((u_{*}^{4}u-u^{5})f'(u)-4u_{*}^{4}f(u)\right)-4u^{5} }{2 r_{h}u^{4}u_{*}^{4}}.
\end{equation}
For the volume enclosed by $RT$ surface we use the invariant quantities of Ricci scalar, $R$, and Krietschmann, $R_{\mu \nu \rho \sigma}R^{\mu \nu \rho \sigma}$. Based on Relations (\ref{form505}), (\ref{form50105}), and (\ref{form1100}), we, therefore,  have:

\begin{eqnarray}
Invariant{C}[R]&=& \frac{Vol(\Re)}{8 \pi G_{4}}\int_{u_{*}/r_{h}}^{\varepsilon /r_{h}}R\frac{x(u)du}{u^{3}\sqrt{f(u)}},\nonumber\\
Invariant{C}[R_{\mu \nu \rho \sigma}R^{\mu \nu \rho \sigma}]&=&\frac{Vol(\Re)}{8 \pi G_{4}}\int_{u_{*}/r_{h}}^{\varepsilon /r_{h}}R_{\mu \nu \rho \sigma}R^{\mu \nu \rho \sigma}\frac{x(u)du}{u^{3}\sqrt{f(u)}}.
\end{eqnarray}
where, $R$, and $R_{\mu \nu \rho \sigma}R^{\mu \nu \rho \sigma}$ can be obtained as follows:
\begin{eqnarray}
R&=&-2uf'(u)+6f(u),\nonumber\\
R_{\mu \nu \rho \sigma}R^{\mu \nu \rho \sigma}&=&2u^{2}f'^{2}(u)-8uf'(u)f(u)+12f^{2}(u).
\end{eqnarray}

Now, these topological invariant quantities may be plotted as a function  of temperature to explore the behaviour of metal/superconductor phase transition. It is interesting to note that the invariant quantities are able to identify the phase transition in the metal/superconductor system and that the order of the phase transition can also appear for these quantities. As can be seen in all the plots in Figs. (\ref{fig1}), (\ref{fig22}), there is a discontinuous slope at the critical point that presents the second order phase transition. As the backreaction parameter $k$ increases, we are farther away from the probe limit and the discontinuity slope in the critical point is sharp. This is while increasing parameter $z$ leads to a smoother discontinuity slope at the critical temperature. In the superconductor phase, however, all the quantities increase with increasing $\beta$. In the metal phase , in contrast, an inverse behaviour is observed and the quantities drop with increasing $\beta$.

\section{Condensation and Conductivity}\label{cap32}
This section is dedicated to the numerical investigation of the condensation and conductivity \cite{reff29999}-\cite{reff155}. For this purpose, we choose the Lagrangian (\ref{form2}) and the metric (\ref{form4}). To obtain conductivity, we need to turn on a small perturbation $\delta A_{x}=A_{x}(r)e^{-i\omega t}$ in the gauge field in the bulk theory. Therefore, the linearised equation of the perturbation $A_{x}(r)$ is as follows:
\begin{eqnarray}\label{form455}
A''_{x}(r)&+&\left(\frac{f'(r)}{f(r)}-\frac{\chi'(r)}{2}+\frac{z+1}{r}(1-\frac{\beta^{2}e^{\chi(r)}\varphi'^{2}(r)}{r^{2-2z}})+\frac{4\beta^{2}e^{\chi(r)}\varphi'(r)\varphi(r)\psi^{2}(r)}{r^{2z}f(r)} \right) A'_{x}(r)\nonumber\\
&+&\left(\frac{\omega^{2}e^{\chi(r)}}{r^{4z}f(r)^{2}}-\frac{2\psi^{2}(r)}{r^{2}f(r)}(1-r^{2-2z}\beta^{2}\varphi'^{2}(r)e^{\chi(r)}) \right) A_{x}(r)=0
\end{eqnarray}
The behaviour of the scalar field and the approaching of the gauge field to the conformal boundary are as bellow:
\begin{equation}\label{form3003}
\psi(r)=\frac{\psi_{+}}{r^{\Delta_{+}}}+\frac{\psi_{-}}{r^{\Delta_{-}}} \\\ , \\\ \varphi(r)=\mu - (\frac{\rho}{r})^{2-z},
\end{equation}
where, $\rho$ and $\mu$ are the charge density and the chemical potential at the boundary, respectively. $\psi_{+}$ is the source of the dual operator $<O>$. Therefore, the condensation operator is defined as $<O>=\psi_{-}$. At the boundary, we have $\chi \rightarrow 0$ and the asymptotic behaviour for the metric function is as bellow \cite{reff2888}:
\begin{equation}\label{form43}
f(r)=1-\frac{\varepsilon}{2r^{2+z}}+...,
\end{equation}
where, $\varepsilon$ is mapped onto the energy density of the dual field theory in the context of holography.
Using the regularity conditions for the equations of motion (\ref{form6}), (\ref{form7}), (\ref{form9}), and (\ref{form8}), we find $\varphi(r_{h})=f(r_{h})=0$ and $\psi(r_{h}), \chi(r_{h})$ are the constants at the horizon. Therefore, the series expansion of the fields near the event horizon, $r=r_{h}$, are as follows:
\begin{eqnarray}\label{formfff}
\varphi_{h}(r)&=&\varphi_{h1}(r)(r-r_{h})+\varphi_{h2}(r)(r-r_{h})^{2}+...,\nonumber\\
\psi_{h}(r)&=&\psi_{h0}+\psi_{h1}(r)(r-r_{h})+\psi_{h2}(r)(r-r_{h})^{2}+...,\nonumber\\
\chi_{h}(r)&=&\chi_{h0}+\chi_{h1}(r)(r-r_{h})+\chi_{h2}(r)(r-r_{h})^{2}+...,\nonumber\\
f_{h}(r)&=&f_{h1}(r)(r-r_{h})+f_{h2}(r)(r-r_{h})^{2}+... .
\end{eqnarray}
Temperature may be determined using (\ref{form410}) by applying the expansions of the fields near the horizon in (\ref{formfff}) into (\ref{form9}):
\begin{equation}\label{form202}
T=\frac{r_{h}}{16\pi}\left((12-2m^{2}\psi_{h0}^{2})e^{-\frac{\chi_{h0}}{2}} -(\frac{\psi_{h1}}{r_{h}})^{2}e^{\frac{\chi_{h0}}{2}}\right).
\end{equation}

\begin{figure*}
\centering
\begin{tabular}{ccc}
\rotatebox{0}{
\includegraphics[width=0.50\textwidth,height=0.18\textheight]{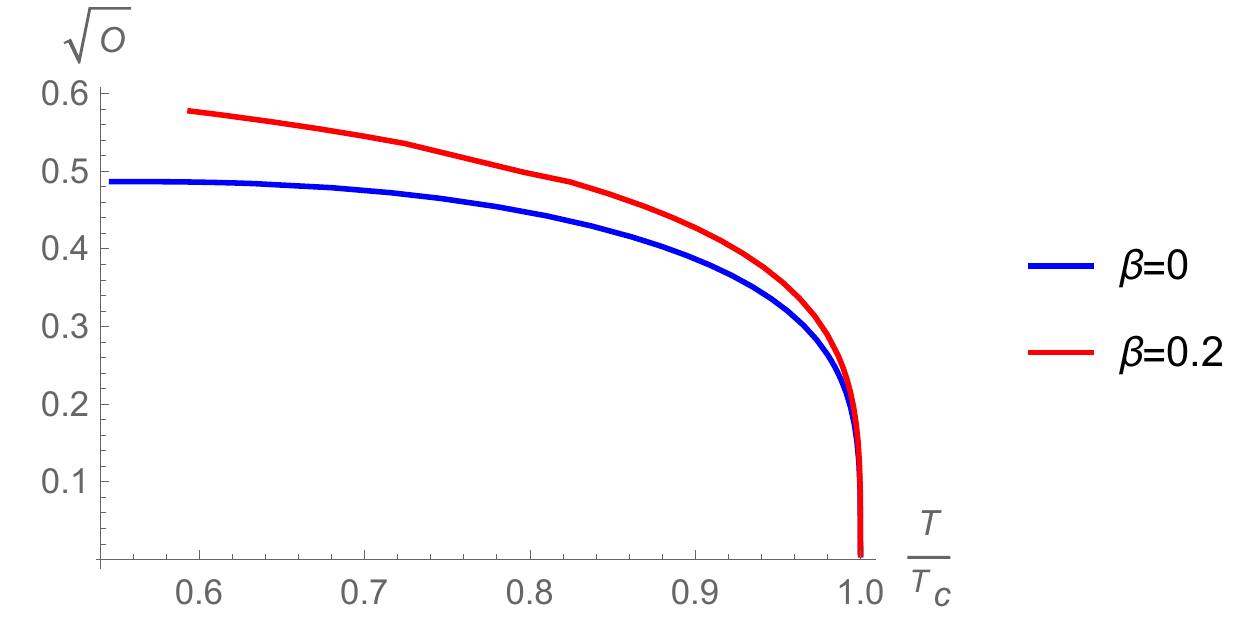}}&
\rotatebox{0}{
\includegraphics[width=0.50\textwidth,height=0.18\textheight]{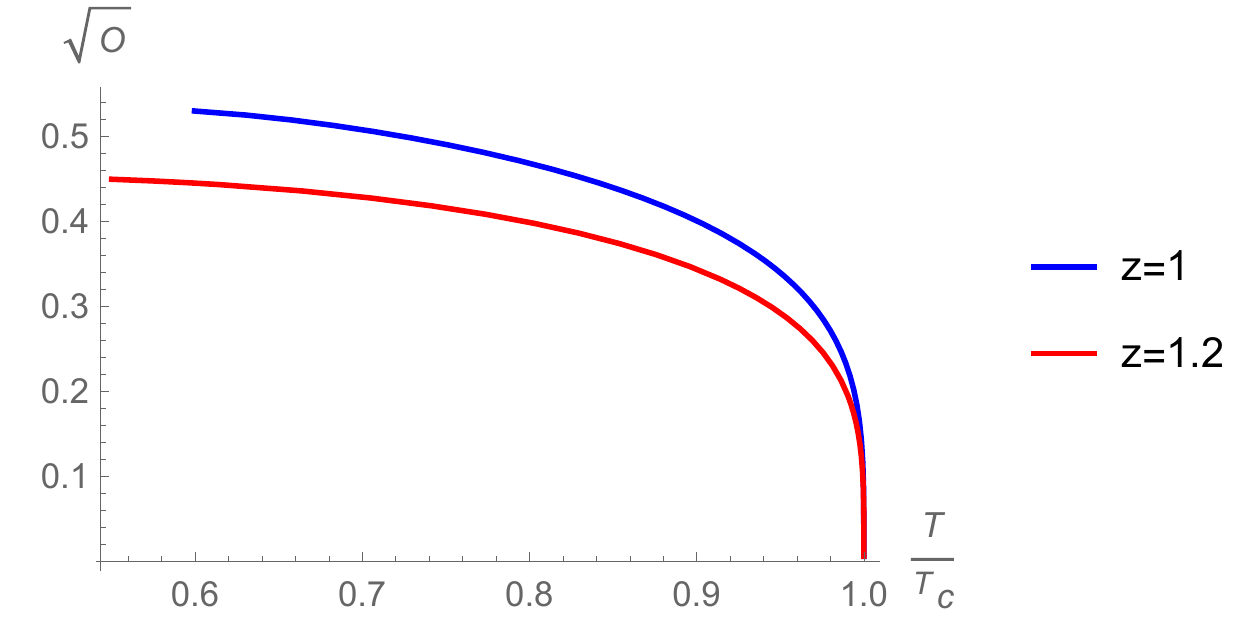}}\\
\end{tabular}
\caption{\label{fig33}The expectation values for the condensation operator for different values of $\beta$  (left) and $z$ (right). In all plots, we use $m^{2}=-2$ and $k^{2}=\frac{1}{2}$. The left diagram is plotted for $z=1$ and the right one for $\beta=0$}
\end{figure*}

To obtain the expectation value for the condensation operator, $<O>=\psi_{-}$, the equations of motion (\ref{form6}),  (\ref{form7}), (\ref{form9}), and (\ref{form8}) should be numerically solved by integrating the equations from the event horizon to infinity. Thus, the mentioned boundary conditions near the horizon, (\ref{formfff}), and the boundary for $f(u)$, $\varphi$, $\psi$ and $\chi$ should be employed before the value for $\psi_{-}$ can be determined. In Fig.(\ref{fig33}), the expectation value of the condensation operator is depicted in terms of $\frac{T}{T_{c}}$. We note that, the critical temperature can be obtained from (\ref{form202}) where the condensation operator vanishes or tends to zero ($\psi\rightarrow 0$).

Integrating Eq. (\ref{form455}) from horizon to infinity and applying the boundary conditions introduced above for $f(r)$, $\varphi(r)$, $\psi(r)$ and $\chi(r)$ near the horizon and the boundary, we can solve (\ref{form455}) numerically. Employing the infalling boundary condition at the event horizon, the gauge field $A_{x}(r)$ near the horizon, $r\rightarrow r_{h} $, may be rewritten in the following form \cite{refff}:
\begin{equation}\label{form46}
A_{x}(r)=(r-r_{h})^{-\frac{i\omega}{4\pi T}}\left( C_{1}+C_{2}(r-r_{h})+...\right),
\end{equation}
where, $T$ is the Hawking temperature. $C_{1}$ and $C_{2}$ are constants to be determined using Taylor expansion (\ref{form455}) near the horizon. Also, near the boundary when $r\rightarrow\infty$, Eq. (\ref{form455}) exhibits the following behaviour:
\begin{equation}\label{form47}
A_{x}=A_{x}^{(0)}+\frac{A_{x}^{(z)}}{r^{z}}+...,
\end{equation}
where, $A_{x}^{(0)}$ and $A_{x}^{(z)}$ are constants. In the following we use the on sell action to obtain the two point function of the current operator:
\begin{equation}\label{form101}
S=\int_{r_{h}}^{r_{\infty}}dr \int d^{3}x\sqrt{-g}L.
\end{equation}

Considering the quadratic approximation for the gauge field perturbation turns Eq. (\ref{form101}) into the following form:
\begin{eqnarray}
S&=&\int d^{3}x\int_{r_{h}}^{r_{\infty}}dr \frac{r^{z+1}e^{-\frac{\chi(r)}{2}}}{2}\left\lbrace \left(2\psi^{2}-\frac{\omega^{2}e^{\chi(r)}}{r^{z+1}f(r)}-\frac{\beta^{2}\omega^{2}e^{2\chi(r)}\varphi'^{2}}{r^{4z}f(r)} \right) A_{x}^{2}\right\rbrace \nonumber\\
&+& \frac{r^{z+1}e^{-\frac{\chi(r)}{2}}}{2}\left\lbrace f(r)\left(1+\frac{\beta^{2}e^{2\chi(r)}\varphi'^{2}}{r^{2z-2}} \right) A'_{x}(r)^{2}\right\rbrace .
\end{eqnarray}
Integrating the above integral by parts and using (\ref{form455}), we have:
\begin{equation}\label{form1001}
S=\int d^{3}x\left[\frac{r^{z+1}f(r)}{2}\left(1+\frac{\beta^{2}e^{2\chi(r)}\varphi'^{2}}{r^{2z-2}} \right) A'_{x}(r)A_{x}(r)\right] |_{r=r_{\infty}}.
\end{equation}
Replacing Eqs.(\ref{form3003}), (\ref{form43}), and (\ref{form47}) into (\ref{form1001}), yields following relation:
\begin{equation}
S=\int d^{3}x \left( zA_{x}^{(0)}A_{x}^{(z)}\right).
\end{equation}
According to Ohm's law, a possible relation for electrical conductivity is as follows \cite{reff4,reff2777}:
\begin{equation}\label{form50}
\sigma(\omega)=\frac{<J_{x}>}{E_{x}},
\end{equation}
where, $<J_{x}>$ is a current operator in the boundary field theory
\begin{equation}\label{form51}
<J_{x}>=\frac{\delta S}{\delta A_{x}^{(0)}}=zA_{x}^{(z)},
\end{equation}
and $E_{x}=-\partial_{t}\delta A_{x}$. Thus, the holographic method may be used to obtain electrical conductivity as expressed in (\ref{form52}) below:
\begin{equation}\label{form52}
\sigma(\omega)=-\frac{iz A_{x}^{(z)}}{\omega A_{x}^{(0)}}.
\end{equation}

\begin{figure*}
\centering
\begin{tabular}{ccc}
\rotatebox{0}{
\includegraphics[width=0.50\textwidth,height=0.17\textheight]{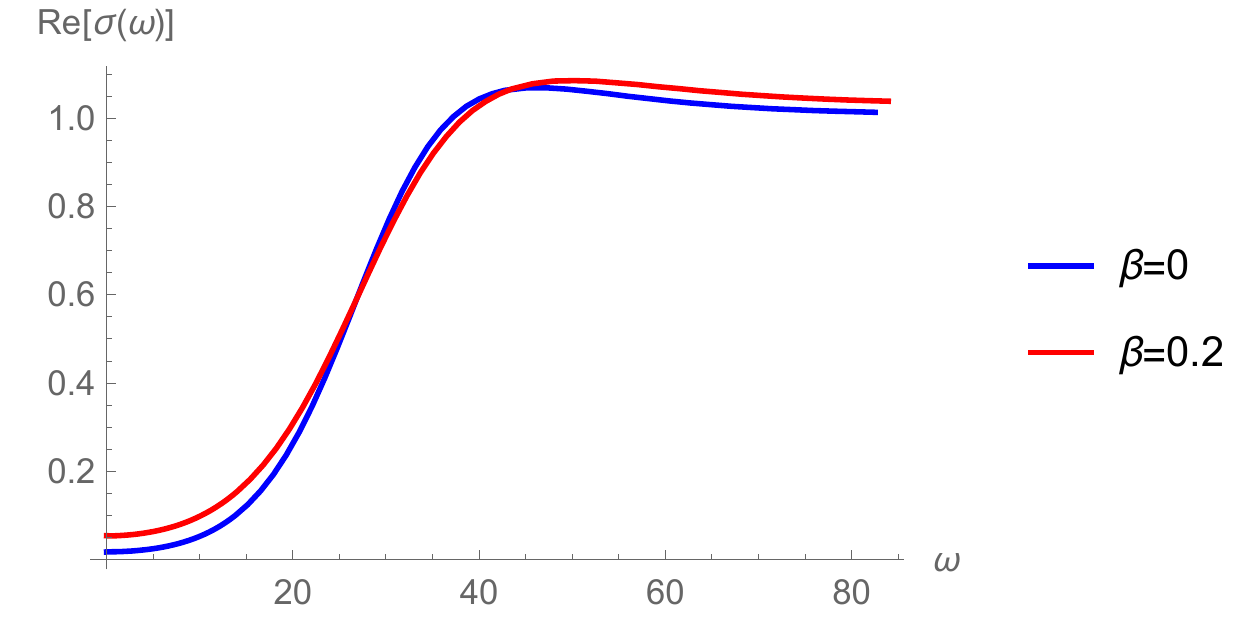}}&
\rotatebox{0}{
\includegraphics[width=0.50\textwidth,height=0.17\textheight]{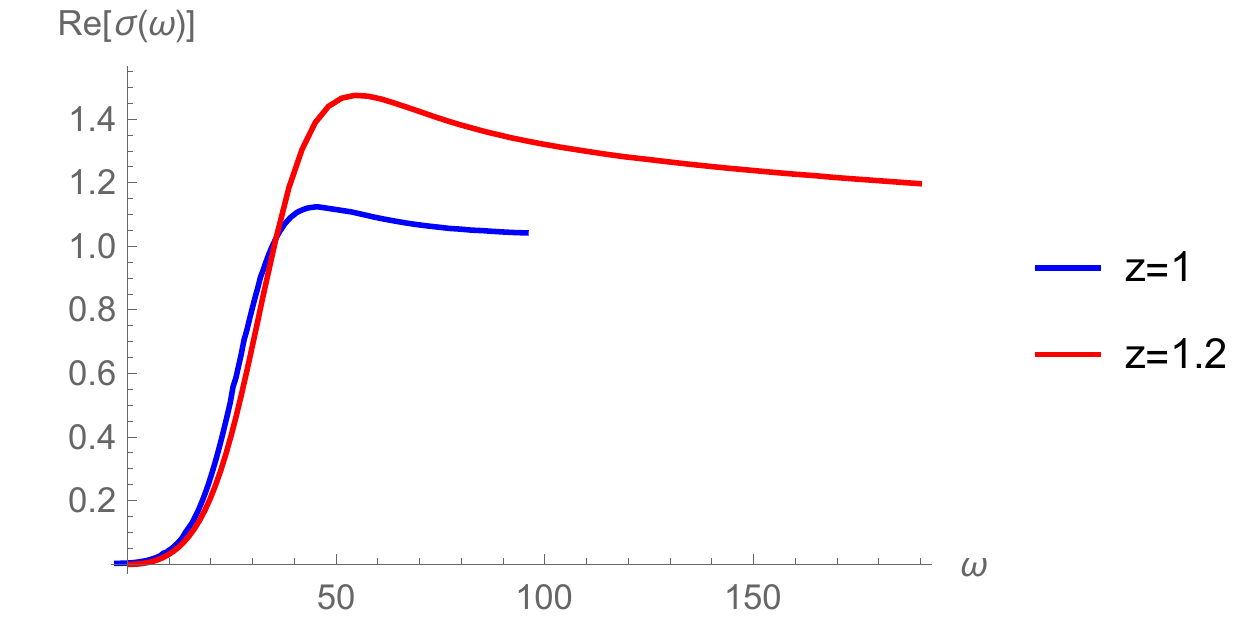}}\\
\rotatebox{0}{
\includegraphics[width=0.50\textwidth,height=0.17\textheight]{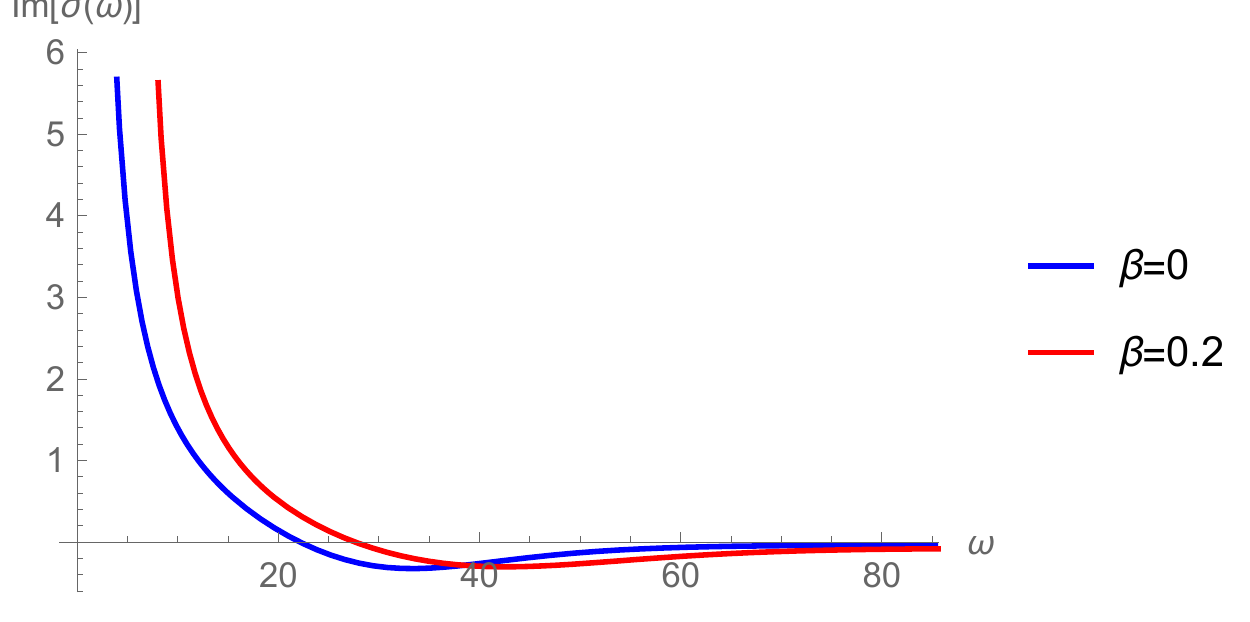}}&
\rotatebox{0}{
\includegraphics[width=0.50\textwidth,height=0.17\textheight]{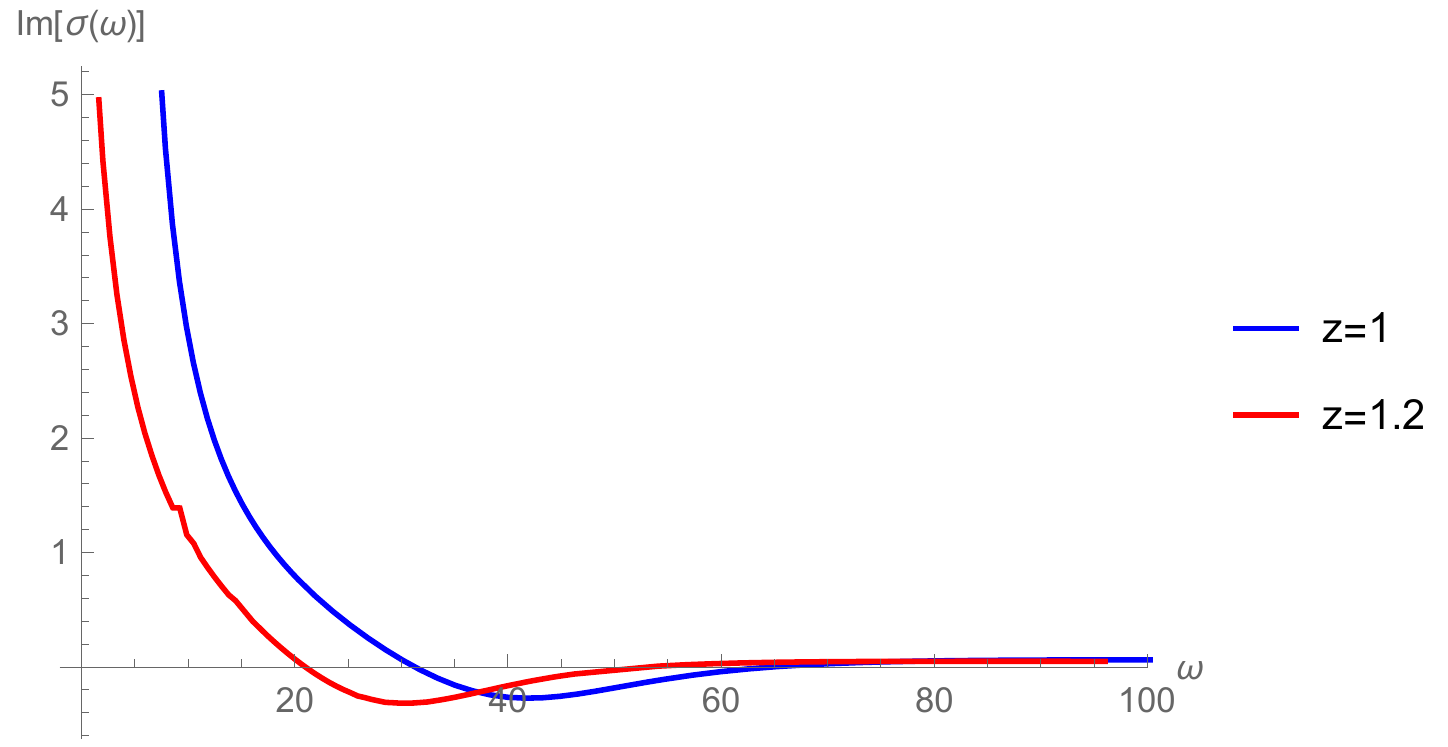}}\\
\end{tabular}
\caption{\label{fig3}The real part of conductivity for different values of $\beta$ (left) and $z$ (right) is presented at the top and the imaginary part at the bottom. In all the plots, $\frac{T}{T_{c}}=0.7$, $\mu=1$ and $m^{2}=-2$. The diagrams on the left are plotted for $z=1$ and those on the right for $\beta=0$ }
\end{figure*}

In Figs.(\ref{fig33}) and (\ref{fig3}), condensation and conductivity are plotted in terms of temperature and frequency $\omega$, respectively. As can be seen in Fig.(\ref{fig33}), the expectation value of the condensation operator increases with increasing $\beta$, but the inverse is observed for the parameter $z$, so that, the expectation value for the condensation operator decreases with increasing $z$. Examination of the diagrams related to conductivity in Fig.(\ref{fig3})  reveals that, for $\frac{T}{T_{c}}=0.7$  and with a fixed value of $z=1$, the real part of conductivity is scaled to one with increasing $\omega$. This is while  the real part of conductivity does not become equal to one at large frequencies with increases in the dynamical exponent, $z$. In this case, the real part of conductivity is enhanced at low frequencies and distances farther away from zero with increasing $z$. These results are consistent with those reported in \cite{reff2077}. It is seen that the imaginary parts of conductivity go to a constant value of zero  at large frequencies.

\section{Conclusion}
We considered a Lifshitz black hole background and a nonlinear electrodynamic to investigate the holographic metal/superconductor phase transition using the topological invariants of the $RT$ surface. We explored the superconductor phase transition by calculating the holographic entanglement entropy and holographic subregion complexity in our model. Also, we used other topological invariants of the $RT$ surface and the volume enclosed by it. It was found that these topological invariants lead to the easier identification of the superconductor phase transition. Topological invariants were obtained for the Lifshitz parameter, $z$; the nonlinear parameter, $\beta$; and the backreaction parameter, $k$. It was observed that these parameters played important roles in identifying the critical points. However, the dictionary for these topological invariants does not exist to obtain their boundary dual quantities. The present work is the first study of these topological invariants to identify the holographic metal/superconductor phase transition points. It will be interesting to use these topological invariants in the holographic QCD and other holographic theories. The dual quantities related to these topological invariants await explanation in future.
Finally, we used the numerical method to study the behaviour of the condensation operator and conductivity in our model.

 \end{document}